\documentclass[journal]{IEEEtran}
\usepackage{graphicx}
\usepackage{epsfig}

\usepackage{mdwmath}
\usepackage{mdwtab}

\begin{document}
%
\title{Fully Three-dimensional Simulation and Modeling of a Dense Plasma Focus}
%
%
%

\author{B.~T.~Meehan,
        J.~H.~J.~Niederhaus
\thanks{B.~T.~Meehan is with National Security Technologies, LLC, a Department of Energy Contractor, e-mail: meehanbt@nv.doe.gov}
\thanks{J.~H.~J.~Niederhaus is a Computer Scientist at Sandia National Laboratories, email: jhniede@sandia.gov}
\thanks{Manuscript received December 1, 2012; revised January 11, 2013.}}

%

\maketitle

\begin{abstract}
A Dense Plasma Focus (DPF) is a pulsed-power machine that electromagnetically
accelerates and cylindrically compresses a shocked plasma in a Z-pinch. The
pinch results in a brief ($\sim$100 nanosecond) pulse of X-rays, and, for some
working gases, also a pulse of neutrons. A great deal of experimental research
has been done into the physics of DPF reactions, and there exist mathematical
models describing its behavior during the different time phases of the
reaction. Two of the phases, known as the inverse pinch and the rundown,
are approximately governed by magnetohydrodynamics, and there are a number of
well-established codes for simulating these phases in two dimensions or in
three dimensions under the assumption of axial symmetry.  There has been little
success, however, in developing fully three-dimensional simulations. In this
work we present three-dimensional simulations of DPF reactions and demonstrate
that 3D simulations predict qualitatively and quantitatively different behavior
than their 2D counterparts. One of the most important quantities to predict is
the time duration between the formation of the gas shock and Z-pinch, and the
3D simulations more faithfully represent experimental results for this time
duration and are essential for accurate prediction of future experiments.
\end{abstract}

\begin{IEEEkeywords}
Dense Plasma Focus, Magnetohydrodynamics, Simulation and Modeling, Controlled Fusion
\end{IEEEkeywords}

\IEEEpeerreviewmaketitle

\section{Introduction}\label{sec:intro}
%
%
%
%
\IEEEPARstart{T}{here} is an extensive literature on experiments performed with
Dense Plasma Focus (DPF) machines, exploring both fundamental Z-pinch physics
\cite{Mather1964,Mather1965,Filipov} and applications of the DPF to fields like
X-ray lithography \cite{Lee2000}, fusion energy \cite{Lerner}, and modeling of
astrophysics \cite{Pouzo}, to name a few. The vast majority of research has
been driven by theory and experimentation, but there is a shift towards
developing new experiments informed by computational models and simulations.
Many of the simulations that have been used to design DPF experiments are
1-dimensional, in the sense that current, temperature, or expected neutron
yield are computed as a function of a single parameter being varied
\cite{Gonzalez}. There has also been work to develop codes that are capable of
full magnetohydrodynamic (MHD) modeling of the inverse pinch and run-down
phases of the DPF reaction on spatial domains (see Section \ref{sec:setup} for
descriptions of the reaction phases), and there exist 2D simulations modeling
the physics of the DPF, which are often extended to 3D assuming axial symmetry
\cite{Kueny}. Despite some success with 1D and 2D simulations, fully 3D
simulations of the MHD phases of the DPF -- which do not impose symmetry on the
physics of the reaction -- are difficult, and the literature presenting the
results of such simulations is sparse. There are a number of challenges in the
3D modeling, including the computational complexity of the problem, a need for
appropriate initial conditions to ignite the inverse pinch, insufficient
equation of state (EOS) data for the working gas and the electrodes, and
incorporating radiative effects into the MHD model. 

In a DPF, a working gas is charged, forming a plasma, that travels down an
anode surrounded by a cathode, and the speed that the plasma shock travels down
the anode is determined by the initial pressure and voltage in the system. The
Z-pinch occurs when the plasma shock gets to the end of the anode. The energy
available to the Z-pinch is maximized when the current through the DPF is
maximized, which means that the rundown time is one of the most important
quantities to properly simulate. If simulations correctly predict rundown for a
given initial pressure and voltage configuration, those settings can be used in
actual experimentation ensuring that the Z-pinch occurs with maximum energy,
which in turn results in maximum neutron yield. The neutron yield can be
determined experimentally using a Beryllium activation detector \cite{Murphy}
(for deuterium fusion), or a Praseodymium activation detector \cite{Meehan}
(for deuterium-tritium fusion).  Further, when deuterium, or deuterium-tritium
mixtures are used as a fill gas, the neutron yield has a power-law relationship
\cite{Zucker} with the maximum current, which means that it is also important
to be able to faithfully simulate the maximum current. For these reasons, most
of our analysis centers on comparing experimentally-measured current waveforms
to the simulated current waveform.

In this work, we present simulation results of a fully 3D MHD model of a DPF
using the ALEGRA \cite{Rider} multiphysics code developed at Sandia National
Laboratories. The simulations are run in both 2D and 3D, and the results are
compared to each other as well as to experiments run in the DPF lab at National
Security Technologies, LLC. The predictions of the 2D and 3D computations are
qualitatively and quantitatively different, as the 2D simulations show
systematically lower inductance, which results in systematically higher
currents but unrealistically fast rundown times. The 3D simulations predict
lower maximum current values but accurately represent the true rundown time
shown in experimental data. This demonstrates that, despite the symmetric
geometry of the machine, there are three-dimensional effects present in the MHD
physics that must be accounted for in order to faithfully predict the outcome
of DPF experiments.

\section{DPF Physics and the Experimental Setup}\label{sec:setup}
The DPF used in our experiments, and the geometry of which was modeled in the
simulations, is formed of coaxial electrodes in a rarified deuterium atmosphere
(about 7 Torr). A two-stage Marx capacitor bank is charged, and when
discharged, breaks down the gas, forming a shock and starting the ``inverse
pinch'' phase of the reaction, in which the gas expands outward from the anode
to the cathode bars (see Fig. \ref{fig:dpf}).  Once the gas touches the
cathode, the ``run-down'' phase begins, and the plasma moves up the anode until
it Z-pinches at the top of the anode. These two phases, the inverse pinch and
rundown, are approximately governed by magnetohydrodynamics and are the
components of the reaction that are studied and simulated.

\subsection{DPF Geometry and Setup}
\begin{figure}[ht]
\begin{center}
\includegraphics[width=3in]{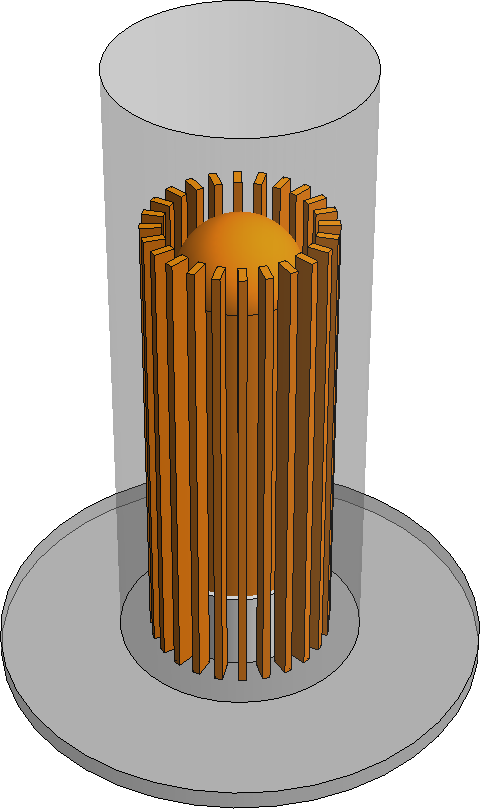}
\caption{A rendering of the DPF used for the models and experiments. The anode
is the dome-topped cylinder in the center; the cathode ``cage'' is the
collection of rectangular bars that surrounds the anode; and the insulator is
visible through the bars, at the bottom of the cathode cage. The vacuum
envelope is represented as the tall cylinder that surrounds the cathode and
anode, and the ground plate is the flat cylinder at the bottom. Some support
features, such as the cathode top support ring, have been left out of the
drawing.}
\label{fig:dpf}
\end{center}
\end{figure}

Fig. \ref{fig:dpf} shows the DPF setup used in our experiments. The outer
electrode (cathode) is formed of 24 copper bars, 0.375 inches thick and 30.75
inches tall, in a ring with an inside diameter of 6 inches. The cathode is at
ground potential, and its bars are shorted at the top with a ring of copper.
The anode is a hollow copper tube with an outer diameter of 4 inches that
stands 23.6 inches above the ground plane, capped with a hemisphere. The vacuum
chamber is 1 foot in diameter, and roughly 6 inches taller than the cathode
cage. A Pyrex insulator tube, which is about 0.5 inches thick and stands 8.63
inches above the cathode base, separates the anode and cathode.

\begin{figure}[ht]
\begin{center}
\includegraphics[width=3.4in]{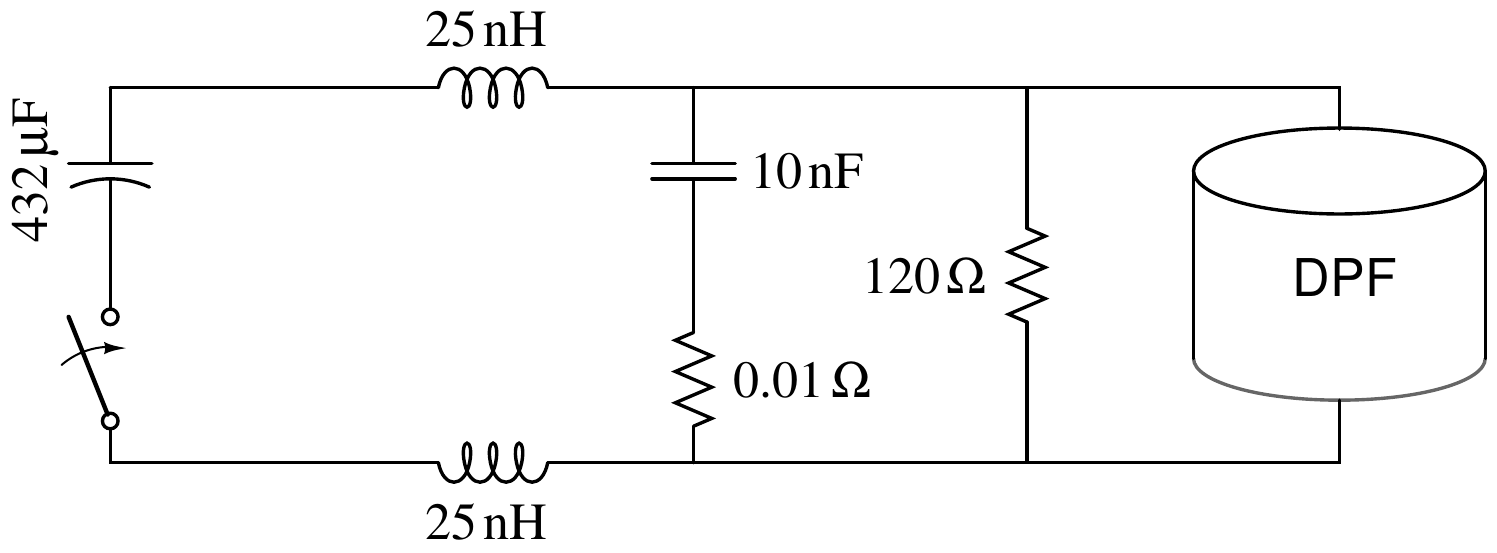}
\caption{The equivalent DPF discharge circuit used in the MHD models. The top
connection of the DPF is to the anode, and the bottom connection is to the
cathode. The switch shown is a collection of eight triggered rail-gap
switches.}
\label{fig:circuit diagram}
\end{center}
\end{figure}

The DPF is driven by a two-stage Marx capacitor bank, which is connected to the
plasma focus tube by 36 coaxial cables. The total capacitance of the bank (when
configured for discharge) is 432 $\mu$F, and the maximum total voltage in
discharge configuration is 70 kV, which makes the maximum stored energy of the
bank 1 MJ. The plasma shock is driven by an external circuit, and the circuit
model used in the MHD simulations is shown in Fig. \ref{fig:circuit diagram}.
The discharge switch in the circuit represents a collection of eight rail-gap
switches that are simultaneously triggered by a single spark gap.  The series
inductance represents the transmission lines that feed power to the plasma
focus tube and was determined empirically by fitting exponentially-dampened
sine waves to the experimental data. The 10 nF capacitor in series with the
small resistor represents the imperfect capacitance of the terminal plates and
the transmission lines that supply the power to the plasma focus tube. The 120
$\Omega$ resistor in parallel with the plasma focus tube is the equivalent
parallel resistance of the safety resistors.

\subsection{Faraday Current Diagnostic}
The discharge current diagnostic is important for comparing simulations to
experiment, because it allows for the measurement of both the maximum current
through and rundown time of the DPF. As was noted above, these are two of the
most important quantities for simulations in order to accurately predict
neutron yield and to ensure that the maximum current runs through the system at
the time of the Z-pinch.  On the NSTec DPF, the discharge current is measured
with a Faraday rotator \cite{Faraday}, which uses the magnetically-induced
linear polarization rotation in quartz fibers to measure the current in a
circuit. The fiber is wound in a circular fashion around the anode, an
orientation that causes the fiber to follow the direction of the magnetic
field, allowing it to accurately measure the current. In (\ref{faraday}), the
polarization rotation angle, $\Theta$ (in radians), is related to the
permeability of the vacuum, $\mu_0$, the current, $I$, the Verdet constant of
the fiber, $V$, and the number of loops the fiber makes around the anode $n$.
The interaction of the magnetic field, $\vec{B}$, with an element of the
fiber's length, $\vec{d\ell}$, is then integrated over the path of the fiber
around the anode.  In our arrangement, the fiber wraps around the anode
$n=5.25$ turns. This path is denoted by $\xi$, and since the fiber is either
parallel with the magnetic field or perpendicular to it, the rotation angle is

\begin{equation}
\Theta = V \int_{\xi} \vec{B} \cdot \vec{d\ell} = \mu_{0} n V I, \label{faraday}
\end{equation}

in MKSA units.  The Faraday rotator current diagnostic is perferred over other
discharge current measurements, such as the Rogowski coil, because the Faraday
rotator gives current measurements that are not dependent on calibration
factors, but rather on an easily measurable geometric quantitiy: the number of
turns around the current to be measured. The only other factor that must be
determined is the Verdet constant for the fiber, which can either be measured
or obtained from a datasheet on the fiber. When properly set up, the Faraday
rotator is a reliable diagnostic.

\begin{figure}[ht]
\begin{center}
\includegraphics[width=3.4in]{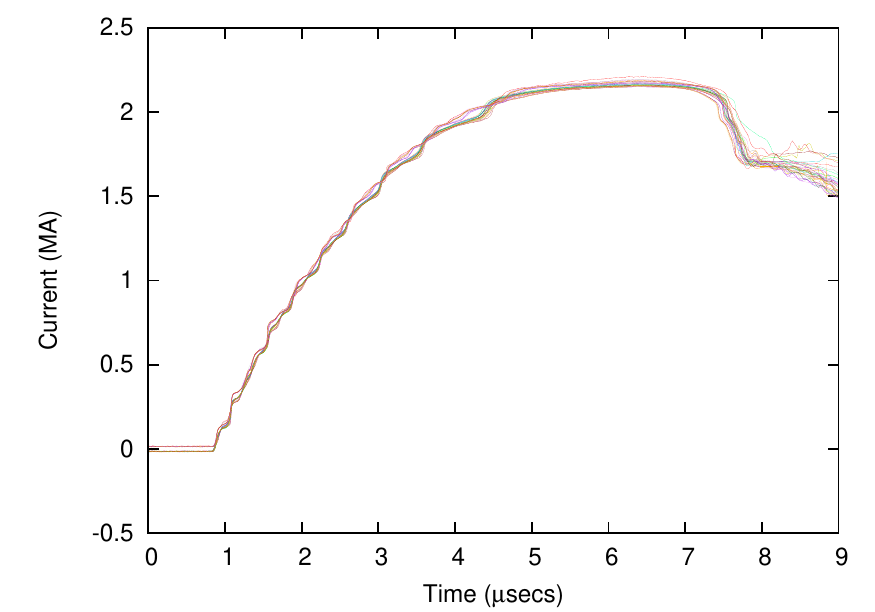}
\caption{Shown is a comparison of the current profiles for thirty-seven DPF
shots, all at the same voltage and pressure (37.5 kV and 7.28 Torr,
respectively). This demonstrates the consistency of the current produced by the
machine, as well as representative Faraday rotator data.}
\label{fig:37_compare}
\end{center}
\end{figure}

\begin{figure}[ht]
\begin{center}
\includegraphics[width=3.4in]{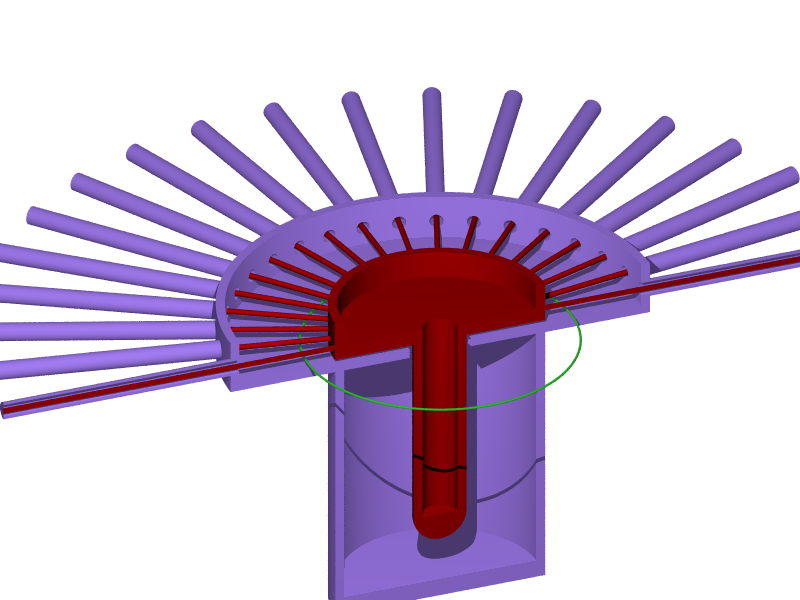}
\caption{Shown is a diagram of where the Faraday coil is placed on the Gemini
DPF. In the cutaway, the red (darker) area represents the parts electrically
connected to the anode, and the violet (lighter) areas are considered to be at
cathode potential. The Faraday loop is shown as a hoop that is under the center
center conductor wires, and above the ground plate.}
\label{fig:faraday setup}
\end{center}
\end{figure}

Fig. \ref{fig:37_compare} shows experimentally-measured results from the
Faraday rotator detector for 37 DPF shots initiated with the same voltage and
pressure. As can be seen, the profiles are all nearly identical, which
demonstrates both the shot-to-shot consistency of the DPF and the reliability
of the Faraday diagnostic. The placement of the Faraday loop is important for
understanding the current that it measures. This is easier to show than it is
to describe, so this is included as fig. \ref{fig:faraday setup}.

\section{Modeling and Simulation of the DPF}
The modeling for this project was performed with Sandia National Laboratories'
ALEGRA-MHD code. ALEGRA is a finite-element, multi-material, arbitrary
Lagrangian-Eulerian (ALE) shock hydrodynamic code designed for parallel
computing. It uses an operator-split edge-element formulation to simulate
resistive MHD in 2D and 3D high-deformation shocked media and pulsed power
systems.

ALEGRA provides fine control over how the simulation is performed through a
text file known as the ``input deck.'' The primary purpose of the input deck
is to define the simulation geometry, material composition, and physics to be
modeled, though it also allows a user to request output and provide runtime
controls. Part of defining the physics of a simulation is setting up the
equations of state for the gases, which was among the most challenging aspects
of the problem. Tabular equations of state provide the most effective means of
modeling the thermodynamic state of material in the simulations, which may be
in the solid, liquid, gaseous, or ionized state. Tabular EOS models were made 
available through ALEGRA's interface to Los Alamos National Laboratory's SESAME 
\cite{Sesame} data. Tabular representations of the Lee-More-Desjarlais (LMD) 
electrical and thermal conductivity model \cite{Desjarlais} were also used 
here. The LMD model combines empirical data with inferences from quantum 
molecular dynamics modeling and density functional theory (QMD-DFT) to provide
a conductivity representation that spans the transition between conducting and 
insulating conditions and has proven quite successful in this ``warm dense 
matter'' regime \cite{Matzen,Lemke}. 

There are two approaches to defining the geometry on which to simulate the
reaction using ALGERA. The first is using ALEGRA's built-in functionality, by
declaring pre-defined volumetric shapes (like spheres, prisms, pyramids, etc.)
and defining the material composition of each shape. For example, many DPF
machines have cylindrical copper cathode bars, and it is possible to define in
the input deck a cylinder made of copper.  The code then generates a 2D or 3D
unstructured mesh that overlays the defined shapes, allowing for mesh elements
to intersect more than one user-defined shape. The second approach to defining
the geometry is by performing the computation on a body-fitted mesh generated
using an external meshing tool. The body-fitted mesh method is a more accurate
method of describing the material in the simulation volume, but is more
time-consuming to set up than the geometric method. ALEGRA's built-in
functionality was used to define the geometry and material composition in all
of the simulations shown in what follows, and boundary conditions are specified
on subsets of nodes or faces within the simulation volume. 

In any simulation, it is important that the geometry and the physics being
simulated reflect the important geometries and physics of the experiment, and
both raise several concerns in our simulations. For
example, it is not necessary to include a faithful model of the vacuum chamber
in our simulations, since the plasma does not interact with the top of the
chamber during the simulation. Evaluating the physics being modeled, it is
important to understand that the plasma in the DPF reaction is driven by an
external electrical network, and a great strength of the ALEGRA-MHD code is
that it has a sophisticated built-in circuit solver that can be used to couple
electrical energy from user-specified circuits into the simulation volume.

Near the end of the simulation, just prior to the Z-pinch, the MHD simulation
begins to become unphysical because of its inability to represent certain
phenomena, such as, the kinetic instabilities which raise the plasma
resistivity. The simulation may run past the point of Z-pinch without crashing,
but the time-evolution of the simulation would be unphysical. Once the MHD
simulation begins to approach the Z-pinch, it is possible to transfer the model
state information to a particle-in-cell model to accurately simulate the
Z-pinch, which has been demonstrated by researchers at LLNL\cite{Schmidt} and
SNL\cite{Kueny}.

Setting up the initial condition for the simulation can also be tricky, since
the DPF's starting state happens when the gas in the chamber breaks down in the
vicinity of the insulating sheath and becomes a ring-shaped plasma shock. The
breakdown of the gas is not covered by MHD physics, so the gas near the
insulator has to be initialized in an artificial state that will quickly
transition to the plasma shock known experimentally to exist in the DPF. We
have found that setting up a thin layer of extremely hot ($\sim 10^6$ K, and
therefore conductive) gas on the surface of the insulator results in the
simulation initiating a plasma shock without causing observable artifacts in
the time evolution of the simulation. Our experience with this initial
condition is that the thin layer of hot gas should be about as thin as the
Pyrex insulator and should touch both the anode and the cathode. Using this
initiation of the plasma shock results in temperatures that match data from
particle-in-cell calculations \cite{Kueny}. The artificially hot gas layer
should stabilize its temperature near the shock temperature, $\sim10^4 K$,
within a few solver timesteps (typically, about 20 nanoseconds).

\subsection{Two-dimensional Modeling}
One and two dimensional models of the DPF are the most common in the
literature, frequently coming in two generic types: empirical models and
finite-element MHD models. The primary strength of empirical models, such as
RADPFV5.5de \cite{Lee2011}, or Scat95 \cite{Tasker} is that they give results
that are often very close to experimental data. A significant drawback is that
they require existing data to fit the model to, and can therefore only be used
to simulate devices that are quite similar to the devices that one has
experimental data for already. The second class are finite-element codes such
as MHRDR \cite{Makhin} and Mach2 \cite{Mach2} that perform MHD modeling in one
or two dimensions or in axis-symmetric 3D. Fully 3D MHD codes are not new, but
can be hard to obtain and are usually more difficult to operate than 2D codes.
Results of fully-3D DPF modeling are not well represented in the literature.

\begin{figure}[ht]
\begin{center}
\includegraphics[width=3.4in]{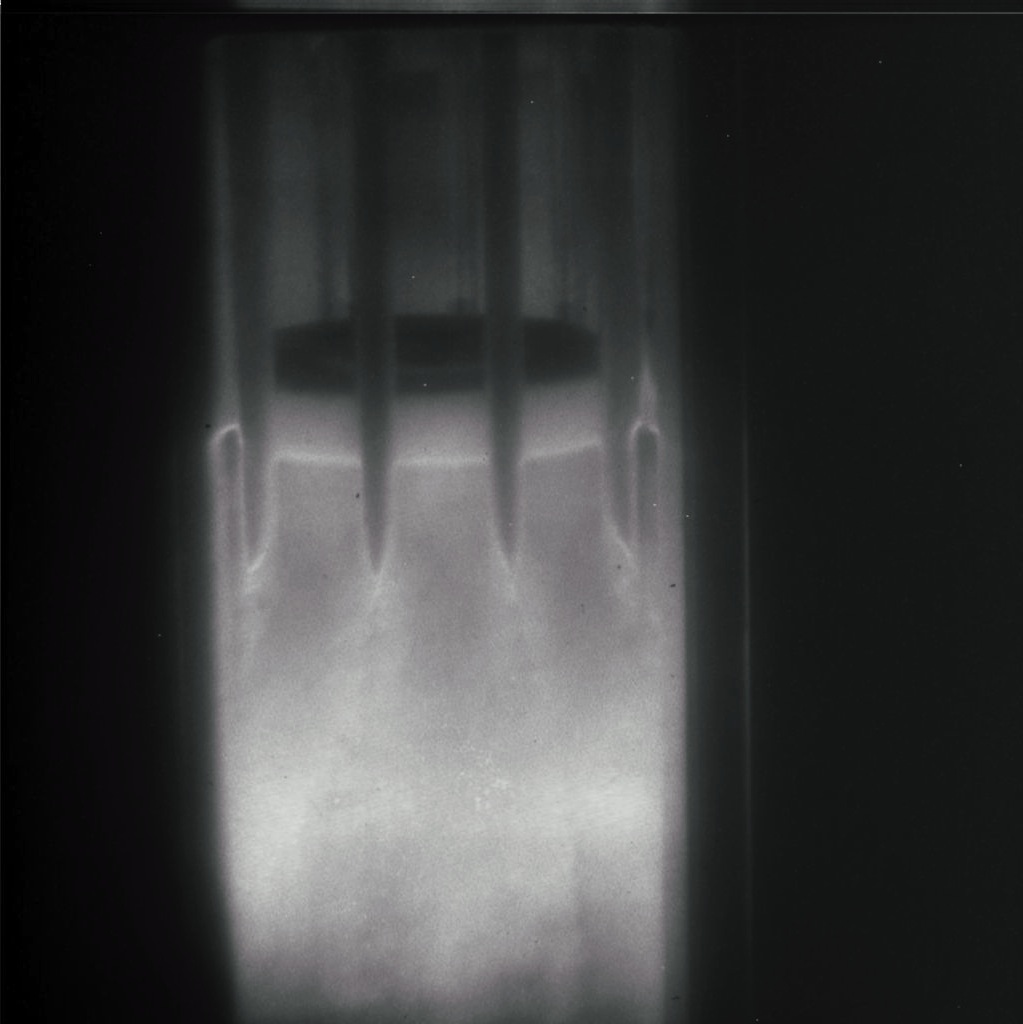}
\caption{Shown is a framing camera picture of the plasma rising up the
electrodes in a DPF. The top of the anode can be seen as the dark disk in the
middle of the bars, which can be seen at the top. The lower region of the
chamber is bright due to the plasma, which has escaped the anode-cathode gap
and surrounded the bars.}
\label{fig:framing camera}
\end{center}
\end{figure}

In many experimental realizations of the DPF, the cathode is a cylinder
composed of metallic bars. While the plasma shock is traveling down the tube
during rundown, the plasma spills outside of the anode-cathode gap through the
spaces between the bars, a phenomenon that can be seen in the framing camera
picture in Fig. \ref{fig:framing camera}. This can be approximated
in 2D simulations, but in practice it is difficult to predict the time
evolution of an actual DPF using these approximations.

Simulations in 2D also impose other geometric constraints: the current density
must either be completely in the radial-axial plane, or perpendicular to it,
but not both at the same time. This precludes modeling situations that may have
currents flowing helically, or situations in which the current may be flowing
asymmetrically or off-axis. These restrictions on 2D simulations are often not
of concern to investigators, who may want to ensure symmetry and simplicity in
their experiments in order to simplify the data they collect. Nonetheless, the
most general, physically realizable simulations are essential for complete
understanding. 

\begin{figure}[ht]
\begin{center}
\includegraphics[width=4in]{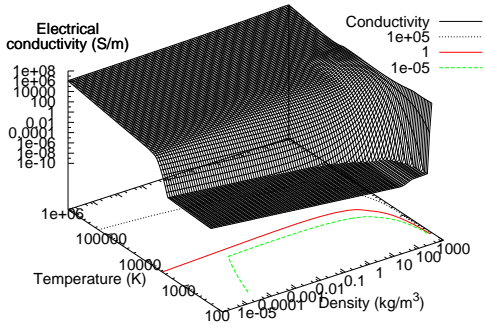}
\caption{The Lee-More-Desjarlais (LMD) electrical conductivity model
for deuterium.} 
\label{fig:LMD}
\end{center}
\end{figure}

The 2D simulation that was run in ALEGRA made the \textit{ad hoc} assumption
that there was a lower density floor below which the material was assumed to
have no electrical or thermal conductivity. The floor was set at a density of
2.5$\times$10$^{-4}$ kg/m$^3$, which was was necessary in order to eliminate
unphysical behavior in the simulation. The LMD model for deuterium plasma,
shown in Figure \ref{fig:LMD}, attributes moderate electrical conductivity to
the plasma at this density for temperatures higher than approximately 1 eV, and
this behavior is suppressed here, in order to cause the plasma shock to travel
properly down the anode of the DPF. 

\subsection{Three-dimensional Modeling}
Fully 3D magnetohydrodynamic codes are available, such as ALEGRA and NIMROD
\cite{Nimrod}, among others, and the benefit of using these codes is that the
electrode geometry in the DPF can be modeled and simulated. The ability to
investigate the effects of electrode structures without presupposing symmetries
allows investigators to gain deeper understanding into how the plasma shock
evolves over time.  The primary drawback of 3D MHD modeling is the
computational complexity of solving the MHD equations on large meshes,
requiring computer clusters to perform simulations in a reasonable amount of
time. Similar to the 2D modeling, the 3D modeling in ALEGRA required the
imposition of a density floor at 2.5$\times$10$^{-4}$ kg/m$^3$ in the 
electrical and thermal conductivity models. 

\begin{figure}[ht]
\begin{center}
\includegraphics[width=3.4in]{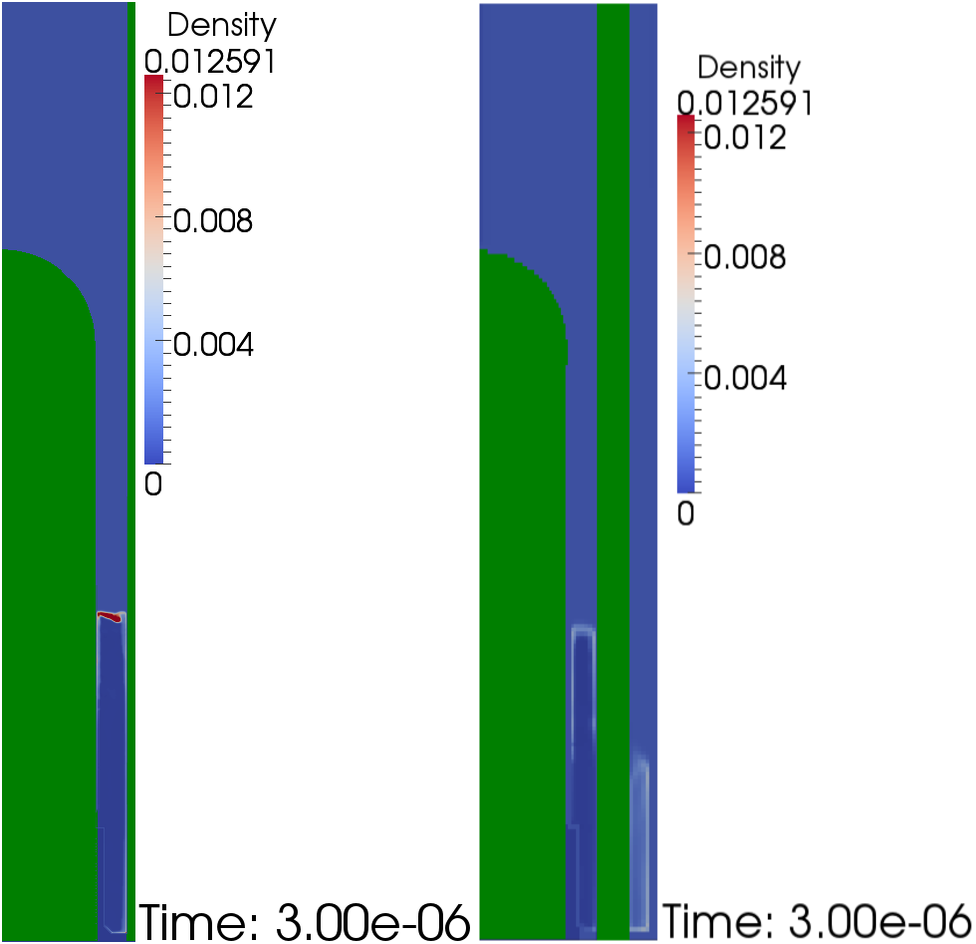}
\caption{This is a comparison of the density on a slice through the simulation
volume at about 3 microseconds. The image on the left shows the 2D simulation
where the plasma cannot flow around the bars, and the image on the right shows
the 3D simulation, where the plasma can flow around the bars. Since the plasma
flows around the bars in the 3D simulation, it also affects the external
impedance as a function of time for the simulation volume.}
\label{fig:density slice}
\end{center}
\end{figure}

Fig. \ref{fig:density slice} shows a representation of the material density
midway through the rundown phase of the DPF in both the 2D and 3D simulations.
The 2D modeling assumes axial symmetry, and the symmetry axis in both
simulation volumes is on the far left-hand side. In the both simulations, the
cathode bars are represented as rectangles on the right of the simulation
domain, and the plasma shock travels from the bottom of the image to the top of
the image, where it Z-pinches slightly above the hemispherical anode top. The
plasma cannot flow around the cathode bars in the 2D simulation as it can in
the 3D simulation. The 3D simulation simulates the entire gas volume of the
chamber, and it can be seen in Fig. \ref{fig:density slice} that the plasma is
slightly slower and less dense in the 3D simulation than in the 2D simulation.
The larger inductance results in longer rundown times and lower maximum current
as compared to the 2D simulation, if all other system parameters are equal.

\section{Experimental and Predicted Results Comparison}
\begin{figure}[ht]
\begin{center}
\includegraphics[width=3.4in]{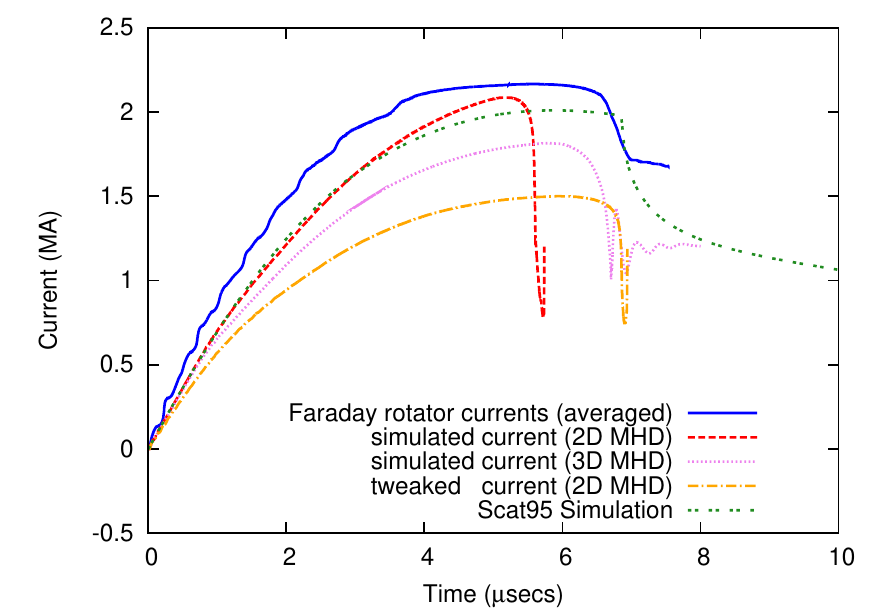}
\caption{Shown is a comparison of the simulated 2D current (red, dashed line)
to the current measured by the Faraday probe (blue, solid line) and to the
simulated 3D current (yellow, dash-dotted line) and a Scat95 simulation (green,
dotted line). The 2D simulation underestimates peak current and severely
underestimates rundown time. The 3D simulation also underestimates peak current
but more faithfully predicts rundown time. Also shown is a 2D simulation (pink,
dotted line) whose input parameters are adjusted to give a rundown time similar
to the experimental data. In order that the 2D simulation match the
experimental rundown time, the peak current is severely underestimated. The
Scat95 simualtion shows better agreement, however, requires iterative
adjustment of parameters to already existing experimental data.}
\label{fig:2d_compare}
\end{center}
\end{figure}

Fig. \ref{fig:2d_compare} shows the current profiles from a 3D simulation, two
2D simulations, and from the Faraday diagnostic of an actual experimental run.
The ``simulated current'' in 2D (red, dashed line) and 3D (yellow, dash-dotted
line) were both initiated using the same voltage and pressure as the
experimental data. The 2D simulation predicts a peak current of 2.08 MA, which
differs from the peak current measured by the Faraday diagnostic (2.17 MA) by
only 4\%. The 3D simulation systematically predicts lower peak currents, due to
the higher inductance of the plasma escaping the cathode bars, and estimates
peak current at 1.82 MA, an error of approximately 16\%. Thus, for estimating
peak current, the 2D simulation is more accurate than the corresponding 3D
simulation.

The more important quantity of interest, however, is the duration time of the
rundown phase of the reaction.  The end of the rundown phase is defined at the
point of maximum derivative of the current profile, which is 6.96 $\mu$sec for
the experimental data. The 3D simulation predicts 6.69 $\mu$sec, an error of
less than 4\%, whereas the 2D simulation predicts a rundown time of 5.59
$\mu$sec, an error of almost 20\%, showing that the 3D simulation vastly
outperforms the 2D simulation in predicting this quantity.

The two primary inputs into the ALEGRA simulation that we have discussed so far
are the initial voltage and pressure, since these are the adjustable initial
conditions of the DPF machine. The simulation allows for other quantities to be
tweaked as well, though, and it is natural to adjust the model parameters to
attempt to match the experimental data more closely. This is difficult with the
3D simulations, since they are too computationally intensive to ``tweak''
parameters one at a time and analyze the changes in output. For the 2D
simulations, however, this is possible, and Fig. \ref{fig:2d_compare} shows the
results of a 2D simulation (``tweaked current,'' pink dotted line) that was
designed to match the rundown time of the experimental results. This required
that the series inductance be adjusted from 25 nH to 28.2 nH. Note that there
is no experimental justification for this change, it is done just to show that
the true rundown time can be achieved with a 2D simulation. This simulation
does not outperform the 3D, however, since the 2D simulation matching the
experimental rundown time results in a far inferior peak current measurement.
Thus a small improvement in rundown time over the 3D gives a large degradation
in peak current. 

Naturally one should use simulation input parameters that represent the
experiment being simulated as faithfully as possible, and the story of Fig.
\ref{fig:2d_compare} is that 2D simulations of a DPF can give quite good
results when the peak current is the quantity of interest. When the rundown
time is of interest, which is more often the case, it is necessary to use the
fully 3D simulation to accurately predict the rundown time of an experiment.
For comparison, a Scat95 simulation was iteratively adjusted to match the
experimental data. While the agreement is good for this simulation, the
parameters used in the match are only good matches for geometries and setups
that are close to this particular case. Otherwise Scat95 achieves results that
are similar to the 2D MHD simulations.

\section{Conclusions}
In this work we have presented results of fully 3D predictive simulations of a
Dense Plasma Focus, using the ALEGRA MHD code from Sandia National Laboratories.
As opposed to 2D or axis-symmetric 3D simulations, the fully 3D models more
faithfully predict the duration of the rundown phase of the DPF, which is
essential for ensuring that the maximum current runs through the system at the
time of Z-pinch, which is required to accurately predict neutron yield. The 2D
simulations are appropriate for predicting the peak current in the DPF, but are
not capable of matching both the peak current and rundown time simultaneously.

\section*{Acknowledgment}
The authors would like to thank Aaron Luttman for helpful comments and
suggestions on the manuscript. We would also like to thank Chris Hagen for
providing support and encouragement for this project, as well as his insight
into the theoretical and experimental operation of the DPF.

Sandia National Laboratories is a multi-program laboratory managed and operated
by Sandia Corporation, a wholly owned subsidiary of Lockheed Martin
Corporation, for the U.S. Department of Energy's National Nuclear Security
Administration under contract DE-AC04-94AL85000.

This manuscript has been authored in part by National Security Technologies, LLC, under
Contract No. DE-AC52-06NA25946 with the U.S. Department of Energy and supported
by the Site-Directed Research and Development Program. The United States
Government retains and the publisher, by accepting the article for publication,
acknowledges that the United States Government retains a non-exclusive,
paid-up, irrevocable, world-wide license to publish or reproduce the published
form of this manuscript, or allow others to do so, for United States Government
purposes.


\begin{IEEEbiographynophoto}{B.~T.~Meehan}
received a B.S. in Physics from the United States Naval Academy in 1995 and an
M.S. in Applied Physics from Stanford University in 1997. Currently he works
with the Dense Plasma Focus Group at National Security Technologies, LLC,
modeling DPFs with magnetohydrodynamics codes.
\end{IEEEbiographynophoto}

\begin{IEEEbiographynophoto}{J.~H.~J.~Niederhaus}
holds a B.S. in Physics from the Virginia Military Institute (2001), an M.S. in
Nuclear Engineering from the Pennsylvania State University (2003), and a Ph.D.
in Engineering Physics from the University of Wisconsin-Madison (2007).  He is
a computer scientist in the Computational Shock and Multiphysics Department at
Sandia National Laboratories. 
\end{IEEEbiographynophoto}

\end{document}